\documentclass{amsart}
\usepackage{amsmath, subcaption, graphicx, enumerate}
\begin{document}

\title[Segmentation \& Workflow for (Digital) Portfolio Maximization]{How \& Why to Use Audience Segmentation\\ to Maximize (Listener) Demand\\ Across a Digital (Music) Portfolio}
\author{Kobi Abayomi}

\maketitle

\begin{abstract}
Digital delivery of songs has radically changed the way people can enjoy music, the sort of music available for listening, and the manner by which rights holders are compensated for their contributions to songs. Listeners enjoy an unlimited potpourri of sounds, uniquely free of incremental acquisition or switching costs \textit{which have been replaced by subscription or rentier fees}. This regime shift has revealed listening patterns governed by affinity, boredom, attention budget, etc.: instantaneous, dynamic, organic or programmatic song selection. This regime shift in demand availability - with the commensurate translation of revenue implications - deprecates current orthodoxy for content curation.  The impulse to point-of-sale model is insufficient in a regime where demand revenue is proportional to demand affinity and each are strongly dependent \textit{time series} processes. We explore strategies \& implications - which are generalizable to any media rights holding firm - from a prediction \& optimization point of view for two straightforward demand models.
%but not the music industry
\end{abstract}
%Listener segmentation, artist \& sound curation, \& with concordant and discordant market level dynamics.

\noindent

\section{A Simple Dynamic Model for Streaming Song Listening}

This paper focuses on modeling demand for a 'song' - generally a two to five minute musical composition  - consumable via some digital delivery service or \textit{Digital Streaming Provider} (DSP) and strategic, macroscopic, inferences that may be deduced from an elucidation of some assumptions around the demand for those songs. %The model we consider assumptions. 

Nile Rodgers, in an interview about his influence on popular music of the 1970's and 1980's recalled this exchange with Miles Davis:

\begin{quote}
Miles would always ask me to make him a hit like how I did for [David] Bowie. I never took him seriously until he covered [Cyndi Lauper's] `\textit{Time after Time.}' I listened to that track and realized he was serious, and like most artists, wanted as many people to hear him as possible. \cite{Rodgers2021}
 \end{quote}

This paper addresses \textit{macroscopic} dynamics of song listening, via model eludcidation of idiosyncratic, dynamic or \textit{microscopic} listener group-by-listener group differences. That is, we focus on the aggregate demand dynamics of a population, or sub-population, enjoying a 'song' as a function of time, aggregating (if not fully eliding) the individual or group-wise 'utilities' -- here \textit{probability} of listening -- into larger group-wise aggregate demand. 

\subsection{Streaming Demand as a Counting Process}

To begin, but without loss of generality, we consider a 'song' a \textit{de novo} offering: a new, or new version of a composition yielding a demand curve with a fixed point at (0,0): time zero, just as a song is released - or, in the parlance, \textit{dropped}.

In similarity with \cite{Ivaldi2023}, the model for volume of listening, or listener response to a listening 'opportunity', is a counting process where any individual listener enjoys a song with a (not-necessarily) time variant probability $P_{t,i}$\footnote{In this paper capital letters represent random processes, lower case letters observed or observable values. We rely on the random processes as \textit{U-statistics} \cite{Hoeffding} for measurability, and other, assumptions.}:

\begin{equation}
U_{t,i} = \begin{cases} 0 \; w.p. \; P_{t,i}\\
1 \text \; w.p. \; 1-P_{t,i}
\end{cases}
\end{equation}

%We make what we believe ae observable suppositions about these dynamics, which yields a class of time dependent processes to be managed via an optimization scheme.

This models the time dependent aggregated listening (as an \textit{affinity} curve, say -- an aggregation of individual \textit{observed} utilities for listeners) as \textit{demand curves}:

\begin{equation}
Y_t^j= \sum_{i=1}^{N_{t}^{j}} U_{t,i}
\end{equation}

the cumulation of individual listening demand within each \textit{listening strata}, $N_t^j$, yielding a song-level demand curve.

\begin{equation}
Y_t= \sum_{j=1}^{J} \iffalse w_j \fi Y_t^j
\end{equation}

\iffalse with $\mathbf{w} = \{w_j\}_{j=1..J}$; \fi $1,...,J$ indexes the collection of listener strata and may coincide, or not, with the categorically defined DSPs.\footnote{\iffalse It may be desirable that $\mathbf{w}$ is convex on the space -- of time dependent -- demand curves. As well \fi The collection of listening strata $\{N_t^j\}_{j=1,..,J}$ (in an abuse of notation used for both the stratum and its cardinality) are not necessarily disjoint.}
These curves are models for listener preference, over time, for coherent - but not necessarily identical - patterns of listening demand or consumption unique to the 'listening mode.' A 'listening mode' can be as idiosyncratic as the set of rules (or some of the set of rules) which increment a song as a fully 'listened to' stream on a particular DSP (30 seconds of listening, say, on a particular provider), or as collective as merely on which DSP a song is enjoyed. 

Our model is an extension of the well-known `DSP-wise' differences in listening affinity to more general segmentation of listening affinities. One way to convey this is to say that any listener, at any time, may be exposed to (and listen to) any song for any reason - in fact, multiple reasons.

\begin{figure}[htbp]
\begin{center}
\includegraphics[height=7cm]{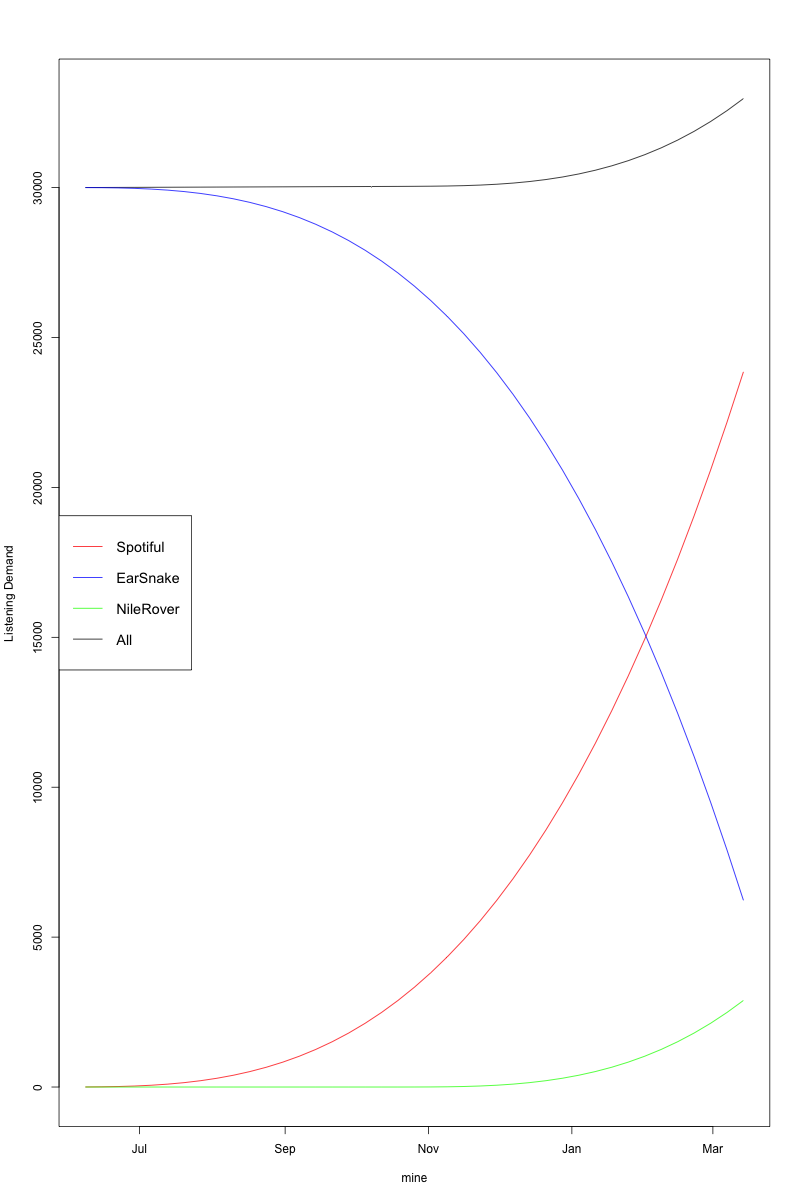}
\caption{Illustration of song demand over time. Each curve, $Y_t^j$, is a illustration of demand for similar audience strata modes. Here the audience strata and the DSP coincide perfectly (though they needn't): the number of people listening to the song on a DSP at a coincident time within an agreed upon time indexing. In this cartoon $J=($Spotiful, EarSnake, NileRover$)$: three made-up DSPs each convey different demand curves. One can imagine an example narrative, constructed from the figure, for example -- \textit{mine} dropped precipitously on EarSnake, built slowly on NileRover and much more quickly on Spotiful -- that explains the different demand curve shapes. Behind each unique demand curve there is differential performance of the song over time and thus differential listening affinity among the population strata which yield each curve. Here $|N_t| =| \cup_{j} N_t^j | = \sum_j |N_t^j |$ - though that needn't be the case as each or any listener can fall to each, any, or multiple strata.} 
\label{fig1}
\end{center}
\end{figure}

In Figure \ref{fig1} a cartoon graphic of listening demand for a song for a 40 week interval: the height of each curve is the number of listeners within each week on each DSP, say; each colored curve aggregates listeners for a unique subscription service within each week. The black curve is the total from each and the overall demand curve. This illustration should be familiar to music industry executives and/or artists: an important heuristic for modeling song performance is that \textit{it should be clear that a song performs differentially} (over time) on different platforms. DSPs can appeal to different audiences, with possibly different listening preferences; each DSP may offer variegated subscription plans, which may appeal to listening preferences heterogenously. 

The aggregate curve in Figure \ref{fig1} -- in black at the top -- conveys a slow steady growth in listening demand. The other curves, on audience (sub) segments illustrate the differential listener affinities (at least, on different DSPs). This sort of rich, differential, picture of demand that is invaluable to a modern content rights holder.

\subsection{Other Counting Process for Streaming Demand}

Content rights holders typically receive intermediated information on listener demand, via the DSPs, in a way that is similar to data scientists in advertising technology. To account for this 'schmutzdecke'\footnote{From my past as an environmental statistician} we augment the naive observed data models with processes offered by latent or hidden features. Let

\begin{equation}
Y_t^{+} = \bigwedge_{N_t^1,...,N_t^J} \sum_{i=1}^{N_t^j} U_{t,i}
\end{equation}

a boundary process, on the best possible audience strata -- i.e. with maximum listening affinity. And let

\begin{equation}
Y_t^{-} = \bigvee_{N_t^1,...,N_t^J} \sum_{i=1}^{N_t^j} U_{t,i}
\end{equation}

be the lower boundary. 

Content rights holders are concerned with song performance -- \textit{and the ability to characterize a song's performance} -- in the presence of confounding factors: temporality, ambient head or tailwinds, DSP idiosyncrasy, bad luck, etc. There are many hard to quantify explanations for song performance. Fixing $Y^+$ and $Y^-$ as the extremal demand processes, \textit{with respect to the process model}, can yields stable comparative models for performance characteristics.

\begin{figure}[htbp]
\begin{center}
\includegraphics[height=7cm]{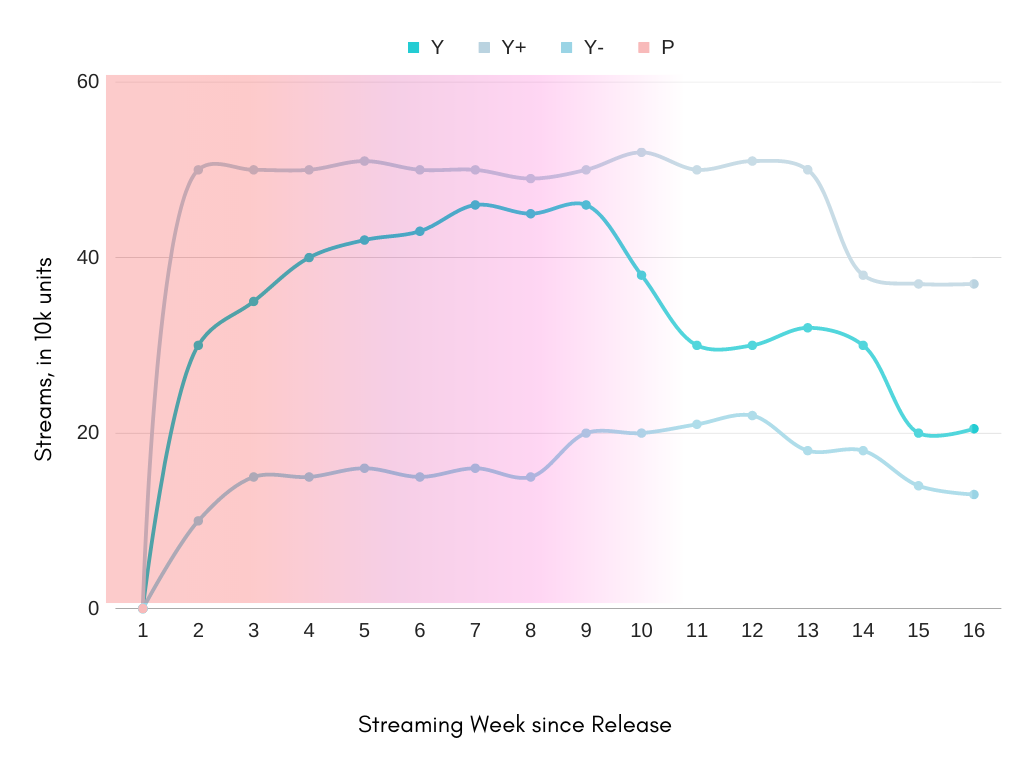}
\caption{Illustration of processes for song demand over time. The curves -- the max value process, the observed demand curve and the minimum value process -- are envelopes for the expected demand over time. Here, the graph is shaded by the 'temperature' of the underlying aggregate affinity process $P_{t,i}$. Affinity for the song begins to 'cool' in week 8.}
\label{fig2}
\end{center}
\end{figure}

\subsection{Model for Listener Affinity}

The model for listener affinity is 

%\begin{equation}
\begin{align}\label{eq:listeneraffinity}
P_{t,i \in j} &= \boldsymbol\theta^j \mathbf{x}_{t} + \boldsymbol\gamma^j \mathbf{z}_{t}\\
U_{t,i \in j} &\sim Ber(\boldsymbol\theta^j \mathbf{x}_{t} + \boldsymbol\gamma^j \mathbf{z}_{t})
\end{align}
%\end{equation}

Remember that the individual listening affinities are collected within listening strata $\{N_j^t\}_{j=1,..,J}$, which are arbitrarily coherent, \textit{but not necessarily disjoint} groups such that 

\begin{equation}
U_{t, i\in j} \sim Ber(\boldsymbol\theta^j \mathbf{x}_{t} + \boldsymbol\gamma^j \mathbf{z}_{t})
\end{equation}

the listening curve can be modeled as from an individual listener or, say, where modeling all the listener demand, at a particular ratecarding at a DSP, as equivalent as from one or a few audience strata. Covariates for exogenous or ambient effects on demand are collected in $\mathbf{z}_{t,j}$; those for endogenous effects (marketing, complementary media, social media, etc.) are collected in $\mathbf{x}_{t,j}$. Assume the $C$ and $D$ dimensional covariates are non-negative such that: $\mathbf{x} \in [0,1]^{C}, \mathbf{z} \in [0,1]^{D}$. 
 
\subsection{Song Demand via Listening Mode}

\begin{figure}[htbp]
\begin{center}
\includegraphics[width=11cm, height=7cm]{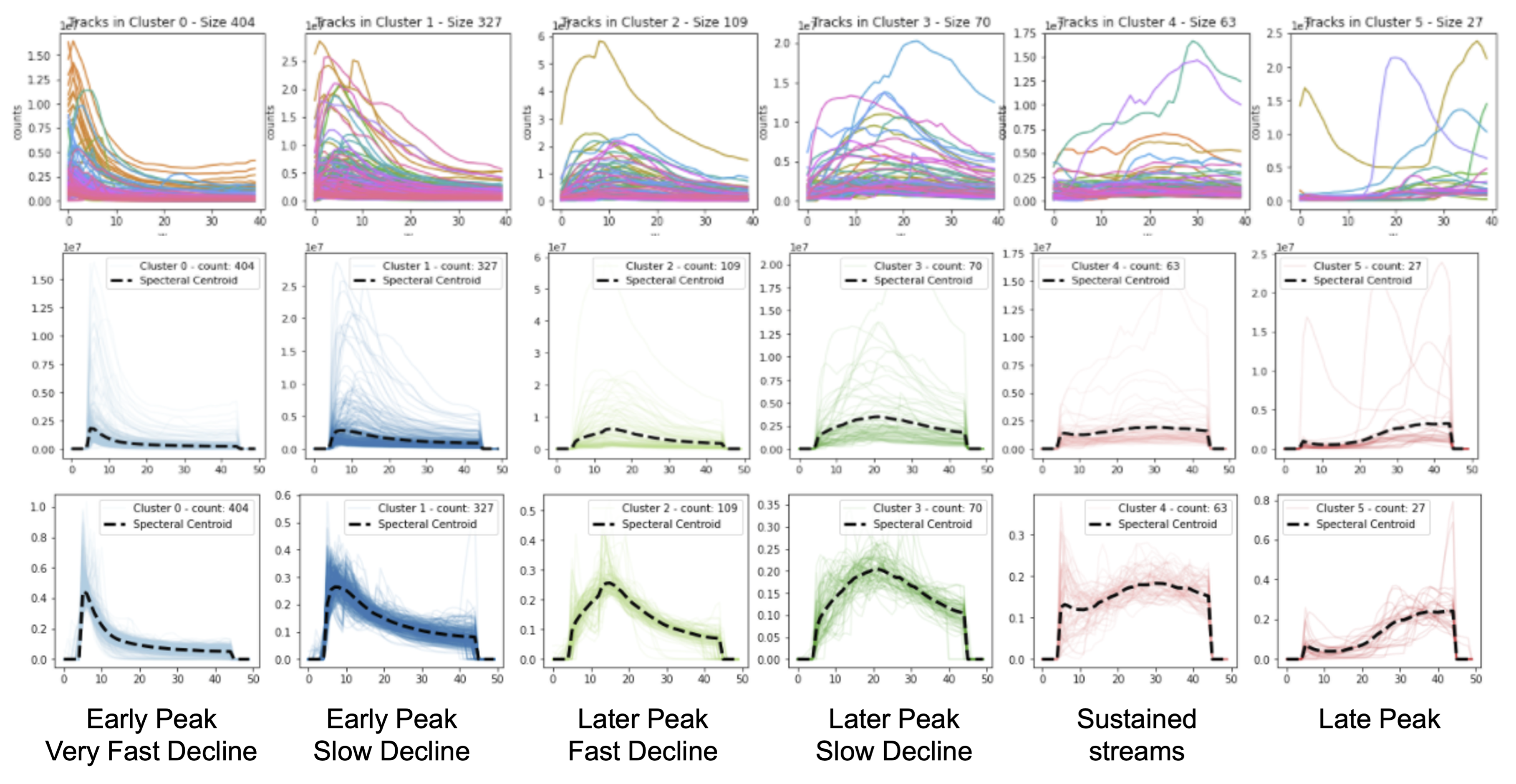}
\caption{Illustration of \textit{modes} of song demand, from observed song demand on a popular streaming service in calendar year 2021. Each \textit{de novo} demand curve was translated to $(0,0)$, i.e. release date vs. zero number of listeners to start. Time is incremented in weeks. As processes, each curve (type) traces the number of listeners in each week.  Successful partitioning of listener types can yield empirically disjoint or differentiable curve types \cite{FPetitjean2011}. Each column are categories of types of listening demand curves is a illustration of demand curves on similar listening modes - the number of people listening to the song on, say, Spotify, at a coincident time within an agreed upon time indexing.}
\label{fig3}
\end{center}
\end{figure}

Figure \ref{fig3} is a plot and characterization of observed demand curves for 1,000 \textit{de novo} songs, with demand curves observed in calendar year 2021, on a popular streaming service. The demand curves were classified by k (=7) mean classification via the Python {\tt tslearn} toolkit to illustrate differences in song demand curves.

Figure 3 points to varied \textit{modes} for listening and song demand: song demand peaks and decays with regular, differentiable characters. Modeling the incidental processes $U_t$ through to the extremal process curves, $Y_t^+, Y_t^-$ lets the model be flexible for the available data granularity. 

\subsection{The importance of audience segmentation}

A feature of this model is to be able to model listening affinity/utility as i.i.d within audience segment. Let audience segments $N_t^1,...N_t^J = \{N_t^j\}_{j=1,..,J}$ be a \textit{covering} s.t

\begin{align}
\begin{split}
N=N_t &= \bigcup_j N_t^j\\
s.t. & \; \forall t \in \{1,...,T\},\\
| \bigcup_j N_t^j | &\leq \sum_j N_t^j 
\end{split}
\end{align}

One can think of an audience segment as a listening group which responds similarly to listening stimuli (at a particular time); within each segment we model the utilities as i.i.d. - random but identically distributed. The $\{N_t^j\}_{j=1,..,J}$ are non-disjoint because individual listeners may occupy more than one utility for listening (at a particular time) a particular song.\footnote{N.B. that the time index for streaming demand modeling can be coarse, where each increment is one week.} The ability to segregate demand to unique audience segments and model differences in effects is important. Let

\begin{equation}
N_t^{\times} = \bigoplus_{j=1}^J N_t^j
\end{equation}

be the `sparse' audience: $N_t^{\times} = N_t - \{N_t^j \bigcap_{j,j^*} N_t^{j^*}\}_{j,j^* \in J}$, with $\oplus$ the symmetric difference operator.

Contemporary work on streaming demand (\cite{Ivaldi2023}, \cite{Ordanini2018}) elides listener level utility with aggregation, perhaps as user level data are hard to come by. The audience segmentation device in this paper joins varied hierarchical level listening demand data with listener level utility models (\cite{Tam2021}, \cite{Wojtowicz2021}). This resonates with the both the spirit of (\cite{Candia2019}) and the similarities in theoretical process models they derive and both they and we observe in data.

\section{Covariate Models for Processes \& Forecasting}

Within any coherent audience segment $i \in j$ the affinity 

\begin{equation}
\hat{\mathbb{P}}(U_{i \in j, t} = 1)= logit^{-1}\{\boldsymbol\theta^j \mathbf{x}_{t, i \in j} + \boldsymbol\gamma^j \mathbf{z}_{t,i \in j}\} = \hat{P}_{t, i \in j, t}
\end{equation}

can exploit models for binary processes: here we can write and use the estimators for the segment-wise affinities via a logistic model. Straightaway the estimators for effects of ambient or ($\boldsymbol\theta$) covariates ($\mathbf{x}$) or the effects ($\boldsymbol\gamma$) of business levers ($\mathbf{z}$) can be modeled using individual, user level data -- if available.  Where these data aren't available -- for example Apple Music's API does not offer granular, user level data --  we can use segment-wise counts and covariates and then we can appeal to natural counting process models, for these aggregates. For example, for observed demand curve $y_t$, for audience segment $j$, the distribution of the size of the audience strata is:

\begin{equation}
\mathbb{P}(N_t^j = n_t) = {n_t -1 \choose y_t - 1} P_{t,i \in j}^{y_t} (1-P_{t, i \in j})^{n_t - y_t}.
\end{equation}

The Negative Binomial distribution relates the demand curves' observed value, $y_t$ to the size of the listening strata $N_t$ \textit{in terms of the covariates} as $P_t$ is covariate dependent. More straightforwardly Poisson or Negative Binomial regression can as well specify the effects of the covariates on the demand curves:

\begin{equation}
log(\mathbb{E}(Y_t | \mathbf{x}_t, \mathbf{y}_t)) = \boldsymbol\theta^j \mathbf{x}_{t} + \boldsymbol\gamma^j \mathbf{z}_{t}\label{eq:logconditionalexpectation}
\end{equation}

and control charts for covariate effects can be easily generated with

\begin{equation}
\mathbb{E}(Y_t | \mathbf{x}_t, \mathbf{z}_t) = e^{\boldsymbol\theta^j \mathbf{x}_{t} + \boldsymbol\gamma^j \mathbf{z}_{t}}\label{eq:exponentialconditionalexpectation}
\end{equation}

as conditional demand curves given proposed ambient or endogenous predictors.

\subsection{Fully Bayesian Workflow for Streaming Demand}

Here it is important to invoke a modeling perquisite: translating the songs to a time-demand interval beginning at $(0,0)$. This condition is met if data for release dates and listening demand beginning from release are available. This condition though is not always necessary, nor it is necessarily sufficient. Consider a model forecasting demand behavior for a song in \textit{deep catalog}: a song that was released many years ago. We illustrated in Figs. 1-3 the \textit{growth-decay} character of listening demand for \textit{de novo} songs; these demand patterns may exist within several alternate or similar periodic behaviors. 

For example, when an audience segment of young listeners discover Stevie Wonder: the mode of growth and decay of listening can be similar, for this strata, to a new release. A forecaster who wants to consider aggregate future demand for a re-release of Stevie's \textit{Jesus Children of America}, say, can't rely fully on only the dynamics of \textit{de novo} songs by comparable artists or even Stevie Wonder himself but \textit{within strata} the assumption is tenable and \textit{across stratum} models are fit on the convolution.

A fully Bayesian setup \cite{Gelman2014} for collecting, training, estimating and updating the model(s) for streaming demand co-ordinates demand response, covariate information and metadata in a framework that is useful for monitoring and gauging song performance in-the-moment and as well yields a full-distributional tableau for a subsequent optimization scheme. 

\begin{figure}[htbp]
\begin{center}
\includegraphics[height=7cm]{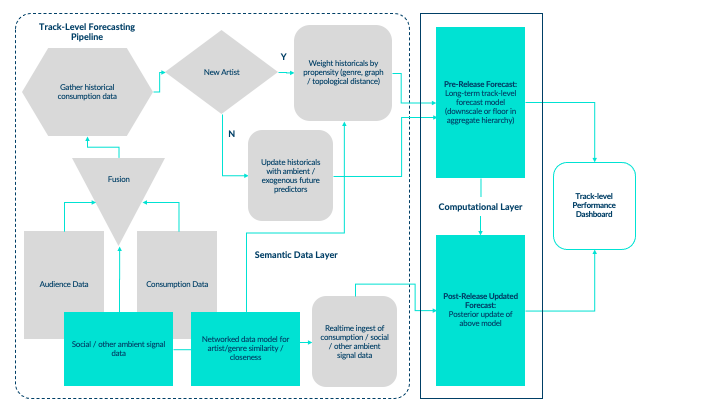}
\caption{Illustration of processes for song demand over time. The dashed box marks processes for data engineering: collation, aggregation, etc. The solid box iterates posterior predictions from the model via updated covariate and marginal posterior information. A dashboard where model priors, empirical processes and posterior inference organizes this information for action and strategy.}
\label{fig2b}
\end{center}
\end{figure}

\begin{figure}[htbp]
Null Model
\begin{align}
\begin{split}
y_t^{j[a]} &\sim NegBin(e^{\boldsymbol\theta^{j[a]} \mathbf{x}_{t,i} + \boldsymbol\gamma^{j[a]} \mathbf{z}_{t,i}}, \omega^{j[a]})\\
\theta &\sim Normal(\boldsymbol\mu_a^x, \boldsymbol\Sigma_a^x)\\
\gamma &\sim TruncNormal(\boldsymbol\mu_a^z, \boldsymbol\Sigma_a^z)\\
\boldsymbol\Sigma_a^x &\sim LkjCorr(\eta_a^x)\\
\boldsymbol\Sigma_a^y &\sim LkjCorr(\eta_a^z)\\
\eta_a^x &\sim \chi^2(\tau^x)\\
\eta_a^y &\sim \chi^2(\tau^z)\\
\omega_j &\sim \Gamma(\alpha_a,\beta_a); \{\alpha_a, \beta_a\}_{a \in A} \; const.
\end{split} 
\end{align}\caption{Bayesian Hierarchical Model for `always on' prediction of streaming demand. Listener stratum are indexed $\{1,...,J\}$ and are `within' artist identifiable. Estimators -- $\boldsymbol\theta \; \text{and} \boldsymbol\gamma$ -- for endogenous and exogenous predictors can enter the \textit{1st} level of the hierarchical model; $a \in A$ collects the `Artist' level of the model. This allows effects on demand to be random across artists. The effects needn't be independent within type: the Lewandowski, Kurowicka, \& Joe \cite{LKJ2009} distribution for the correlation matrices of the effects can be strengthened towards or away from independence by increasing the value of the $\eta$ hyperparameter. The dispersion parameters $\omega^{j[a]}$ can account for difference in observed variation in demand within and across artists. The $TruncNormal$ is the truncated Gaussian distribution, here restricted to the positive reals (see \cite{BarrSherrill}). Further hierarchization -- track to artist to market, say -- follows \cite{Ivaldi2023}. N.B.: only the Negative Binomial distribution arises from first principles of the counting process on utilities; other distributions may be substituted for modeling.}\label{model1}
\end{figure}

Figure (\ref{model1}) elucidates a Bayesian model which captures listener stratum and artist level effects - accounting for differences in utility, say, among the effects on listeners who enjoy only the unique rhythm gbitar, organ and synthesizer on the Ohio Players' single \textit{Ecstasy} and those who have an ear for it in the rest of the album. This hierarchy can of course be extended. Notice that this model is forced by the effect of ambient and planned actions as realized upon listener stratum. These effects in this model are time-invariant and the model itself only accounts for time dependent effects via the value of the predictor processes. This model, thusly, does not necessarily yield a growth-decay process, but for the observed values of the covariate forcings and/or migration out of high affinity (and thus positive estimated effect) listener segments. One can imagine an affinity process where covariate forcing continues at a constant level - especially given the model's partition of listeners into non-disjoint strata. But there is only one Bob Marley.\footnote{Or Michael Jackson, or Jan Hammer, or KraftWerk.}

The model in Figure (\ref{model2}) fixes growth-decay conditions on the listener segment counting processes. In this version of the model the main effects estimators, for the utility forcings, are estimated as projection on subspaces of a phase transition model and in this way mediated or attenuated depending upon the phase of the process. One reason for \textit{not} treating this as a fully Gaussian Process with a Latent Variable (\cite{Li}) is that the generating processes here are only Gaussian in a large numbers regime. Starting from first principles here yields distributional inference even for songs and artists that are less popular, i.e. that stretch the Gaussian assumption on the feature space. The specification of model phase conditional on the estimated change points is equivalent to assuming the main effects estimators within each phase are independent with respect the other phases; that the estimators in each phase are projected \textit{away} from the ancillary subspaces - the other phases (\cite{Lee2019}, \cite{Cook2018}).\footnote{In practice it is desirable to estimate the change points separately and first. The model estimates a subspace for each phase of the process; the effects (ambient and endogenous forcing of the listener segment utilities) estimators are conditional on each estimated subspace. Estimating the phases first can be a common sense check before embarking on the full posterior.} This is addressed this again below.

\section{Conditions for and on Streaming Demand}

Recall that $|N|$ is the total audience available for a song; fix it constant for each time $t$ over the period $\{1,...,T\}$; $T$ usually quite large, each $t$ often a week.. Recall that the $\{N^j_t\}_{0 \leq j \leq J}$ form a \textit{non-disjoint covering} for $N$ s.t. individual listeners $i$ may be in more than one audience segment (at a time) $N^j_t$. The audience segment covering permits differential response to marketing strategies $\mathbf{x_t}$, say, and ambient events $\mathbf{z_t}$ that affect listening affinity -- within each equal time interval $t$ -- via effects $\boldsymbol\theta^{j}$ and $\boldsymbol\gamma^{j}$. Conversationally, the audience segment covering $\{N^j_t\}_{0 \leq j \leq J}$ conveys the \textit{audience segment-wise reason} at a particular time for listening: one time during exercise, another time in an algorithmic playlist of new songs, another time to prepare for sleeping. 

This model places any budget for listening -- from the perspective of the listener -- as a function of the utility curves' $\{U_{t,j}\}_{0 \leq j \leq J}$ response to marketing or ambient impulses $\mathbf{x}_t, \mathbf{z}_t$ -- \textit{i.e. the magnitude of the coefficients $\phi$ and $\psi$} -- and models  incremental listening as membership in a different audience segment (e.g. listeners' ability to listen for a different reason).\footnote{This is an important distinction between the song and utility of listening it at a particular time, for a particular reason. From the perspective of the listener this a model for listening choices; from the perspective of the inventory holder (song creator or curator) its a  model for song demand.}  The impacts of endogenous \& exogenous forcings are conveyed via the individual listening utilities, i.e. \textit{realized probabilities}.

\subsection{Null Model}
Consider the maximization of listening under the null model, where the sole dynamic is listener affinity. From equation (\ref{eq:listeneraffinity}) the user level utility curves are a function of endogenous and exogenous dynamics via effects, respectively $\mathbf{x}_{t,j}, \mathbf{z}_{t,j}; \boldsymbol\theta^j; \boldsymbol\gamma^j$ -- i.e. spend per marketing channel, impulse per social channel, demand per marketing spend and demand per social channel.

Let the endogenous budget $B$ (the amount of money the rights holder has to spend through $T$) \iffalse and ambient stock $D$ \fi for a song be:

\begin{align}\label{eq:budgets}
\begin{split}
B = \sum_t B_t = \sum_t \mathbf{1}^{T} \mathbf{x}_t
\iffalse \sum_{t,c \leq C_m} x_{t,c} \\
D=D_t=\sup_j | \mathbf{z}_j | \\
\fi
\end{split}
\end{align}
 
with $\mathbf{1}$ a vector of ones the same length as $\mathbf{x}$. This is just to say that the rights holder has a finite \& necessarily and wholly exhaustible budget for endogenous forcing.
 
%The amount of ambient or 'social' impulse at time $t$ is maximally bounded by the $[0,1]^{C_s}$ surface.

%\begin{equation}\label
%{sbudgets}
% \infty \geq 
%S = \sum_t S_t = \sum_{t,c \leq C_s} z_{t,c}
%\end{equation}

\begin{figure}[h]
Maximization of Null Model
\begin{align}\label{eq:nullmax}
\begin{split}
\max \mathbb{E}U_{t,i \in j} = \max P_{t, i \in j} = \max_{\mathbf{x}_t \iffalse , \mathbf{z}_t \fi} \boldsymbol\theta^j \mathbf{x}_{t} + \boldsymbol\gamma^j \mathbf{z}_t \\
s.t. \\
\boldsymbol\theta^j \mathbf{x}_{t} + \boldsymbol\gamma^j \mathbf{z}_t & \leq 1\\
\boldsymbol\theta^j \mathbf{x}_{t} + \boldsymbol\gamma^j \mathbf{z}_t  &\geq 0\\
\mathbf{1}^T \mathbf{x}_{t} &\leq B_t\\
\mathbf{1}^T  \mathbf{z}_{t} &\leq S\\
\mathbf{x}_{t} &\geq \mathbf{0}\\
\mathbf{z}_{t} &\geq \mathbf{0}\\
\end{split}
\end{align}\caption{Maximization scheme for Null model. Maximization of the expected utility for any listener, audience-segment-group-($i \in j$)-wise is equivalent to maximizing the probability of listening within segment. The probability term must remain a probability; the budget across channels at a time $t$ is constrained by the total budget available at $t$. Assume that marketing spend and social buzz can only increment.}\label{unforcedmaxscheme}
\end{figure}

A program for the maximization of expected utility for a listener within a particular segment $j$ at time window $t$ is in figure (\ref{eq:nullmax}). Notice that the utility maximization within each listening segment is equivalent to probability maximization within segment. The maximal input for the path, as a function of time, is derived from the Lagrangian for the optimization scheme in (\ref{eq:nullmax}):

\begin{align}
\mathbf{x}^*_{t, i \in j} = \begin{cases} B_t[\boldsymbol\theta^j]^{-1} \; \; where \; \; 0 < B_t \leq (\mathbf{1} - \boldsymbol\gamma^j \mathbf{z})[\boldsymbol\theta^j]^{-1} \\
(1-\boldsymbol\gamma^j \mathbf{z}^t)[\boldsymbol\theta^j]^{-1}  \; \; where \; \; B_t > (1- \boldsymbol\gamma^j \mathbf{z})[\boldsymbol\theta^j]^{-1}
\end{cases}\label{unforcedmax}
\end{align}

where $[\cdot]^{-1}$ is a vector pseudo-inverse. This is to take the maximum of either the scaled available budget $B_t$, or the scaled residue beyond the endogenous effects $\mathbf{z}$; each 'scaled' by the relative effect of endogenous - or business-wise levers - on the song utility, \textit{within each audience segment}. In practice the budget can be reallocated across audience segments - and it should be - to follow the (estimated) effect for greatest gain in audience magnitude.

%\begin{figure}\caption{Illustration of path for maximum utility for a listening segment. The maximal path follow the time-dependent gradient  - i.e. directionally with the component of maximal effect at time t}
%placeholder
%\end{figure}

 %for The segment-wise maximization of probability/utility at each time step vs. 
%The expected number of listeners of type $j$, at time $t$ is:

%\begin{equation}\label{eq:explistener1}
%\mathbb{E}(Y_t^j) = \mathbb{E}(\sum_{i \in j}^{N_j^t} U_{t,i}) = \mathbb{E}(N_j^t) \cdot \mathbb{E}(U_{t,i})
%\end{equation}

%as the individual utilities for each listener are independent even within the same audience segment.

%Conditioning on the size of any particular audience segment yields:

%\begin{align}\label{eq:explistener2}
%\begin{split}
%\mathbb{E}(Y_t^+ | N_t^j ) \geq \mathbb{E}(Y_t^j | N_t^j) \geq \mathbb{E}(Y_t^-| N_t^j)\\
%\Rightarrow \mathbb{E}(Y_t^+ ) \geq \mathbb{E}(Y_t^j | N_t^j) \geq \mathbb{E}(Y_t^-)\\
%\Rightarrow \mathbb{E}(Y_t^+ ) \geq \mathbb{E}(Y_t^j) \geq \mathbb{E}(Y_t^-)
%\end{split}
%\end{align}

\subsection{ADSR/Forcing Model}

\begin{figure}[htbp]
Forced (envelope) Model
\begin{equation}
\begin{split}
y_t^{j[a]} &\sim NegBin(\mathbb{E}(Y(t)), \omega^{j[a]})\\
%\phi_r^{j[a]} &= (\theta^{j[a]})^{-1} \alpha_r^{j[a]}\\
%\psi_r^{j[a]} &= (\gamma^{j[a]})^{-1} \beta_r^{j[a]}\\
\mathbb{E}(Y(t)) & = \alpha_r^{j[a]} + \beta_r^{j[a]} \cdot t\\
%\phi_r^{j[a]} &= Proj_{\alpha_r^{j[a]}} (\theta^{j[a]}) \\
%\psi_r^{j[a]} &= Proj_{\beta_r^{j[a]}} (\gamma^{j[a]}) \\
\theta^a &\sim Normal(\boldsymbol\mu_a^x, \boldsymbol\Sigma_a^x)\\
\gamma^a &\sim TruncNormal(\boldsymbol\mu_a^z, \boldsymbol\Sigma_a^z)\\
\boldsymbol\Sigma_a^x &\sim LkjCorr(\eta_a^x)\\
\boldsymbol\Sigma_a^z &\sim LkjCorr(\eta_a^z)\\
\eta_a^x &\sim \chi^2(u^x)\\
\eta_a^y &\sim \chi^2(u^z)\\
\omega_j &\sim \Gamma(\alpha_a,\beta_a); \{\alpha_a, \beta_a\}_{a \in A} \; const.
\end{split} 
\begin{split}
\alpha_r^{j[a]} = \begin{cases} |\alpha| \geq 0 \; , \; r \leq \tau_A \\
|\alpha| \approx 0 \; , \; \tau_A \leq r \leq \tau_S \\
|\alpha| \leq 0 \; , \; \tau_S \leq r \leq \tau_D \\
|\alpha| \approx 0 \; , \; \tau_D \leq r \leq \tau_R \\
\end{cases}\\ 
\beta_r^{j[a]} = \begin{cases} |\beta| \geq 0 \; , \; r \leq \tau_A \\
|\beta| \approx 0 \; , \; \tau_A \leq r \leq \tau_S \\
|\beta| \leq 0 \; , \; \tau_S \leq r \leq \tau_D \\
|\beta| \approx 0 \; , \; \tau_D \leq r \leq \tau_R \\
\end{cases}\\
\tau_{A} \sim \frac{1}{T-2}\\
\tau_{D,S,R} \sim \frac{1}{T-2} \sum_{t=2}^T \frac{1}{T-t}\\
\end{split}
\end{equation}\caption{Bayesian Hierarchical Model for `always on' prediction of streaming demand with change points and phase shift forcing. Listener stratum are indexed $\{1,...,J\}$ and are `within' artist identifiable. Vector valued estimators for endogenous and exogenous predictors enter the first level of the hierarchy \textit{via the linear equations in equation (\ref{model2max})}. They are still estimated as main effects \textit{per each subspace} of the phase shift model. The phase shift model here has four phases: $A$ attack or growth; $D$ decay; $S$ sustain; $R$ release. The change points for each phase can be estimated simultaneously or before the remainder of the posterior for $y_t$ (here the prior is Restricted Uniform - see \cite{Koop}).}\label{model2}
\end{figure}

The forced model imposes a pattern, or a template of, overarching listening affinity (or song uptake). Refer again to Figure \ref{fig3}. The use of the forcing model is to exploit the regular patterns in aggregate song demand with a model that reduces the inference burden while increasing the explanatory power. Here, we use the \textit{envelope model} -- common to the sound engineering literature as a model for the intensity of a sound over time \cite{Puckett2006}, and a well-known \textit{generative} tool for modifying a sound. Statistically this model is a special case of a \textit{phase transition model} (see \cite{Gomez2018})  - characterized by discontinuities between the phases at the transitions. Referring to the elucidation in equation (\ref{model2}) this model is fit in two steps:

\begin{enumerate}[I]	
	\item \textbf{Fit the change points} The four phases of the ADSR model yield 3 change -- or discontinuity -- points. These can be fit \textit{a priori}, prior to the fully Bayesian estimation of the remainder of the model parameters, or either \textit{a priori} or jointly via the distributional specification in (\ref{model2}) (see \cite{Polunchenko2012}).
	\item \textbf{Fit the partite models} Each phase of the ADSR model is essentially linear: the parameters to be fit are the slopes and intercepts for each linear part; the effects between the endogenous and exogenous covariates; the distributional hyperparameters for dependency between and precision of those effects.  
\end{enumerate} 

The model is conceived to capture dynamics for \textit{de novo} songs - songs new to an audience of listeners,\footnote{To borrow jargon from advertising technology, the \textit{in-flight} period for an advertisement is the length of time an advert is placed within media for \textit{impressions}.} yet is flexible to serve for songs with varied observed release times and listener exposure.  

In the forcing model the endogenous and exogenous effects are estimated jointly with the partite linear model parameters. This is simply to say that the model flexibly estimates the effect on listener affinity within audience segment and subject to the growth/decay phase of the song, given the ADSR model. 

The equations in (\ref{model2}) \& (\ref{model2max}) now specify a Bayesian hierarchy similar to the unforced model but with estimators for effects $\mathbf{\theta}, \mathbf{\gamma}$ that are constant within phase. This simplifies the maximization scheme. For example, in phase $[I]$ the maximum expectation is at time $t_A$, within this phase the estimating equations for effect are $\alpha = 0$ \& $\beta=\frac{\mu_{t_A}}{t_A}$. The mean value function in this phase, $\mu_{t_A}$ is defined as in the unforced model.

% https://www.youtube.com/playlist?list=PL86D5A3CA4C8BF2E8
% https://www.dsprelated.com/	
% https://www.dsprelated.com/freebooks/mdft/
% https://www.dspguide.com/	

\begin{figure}[htbp]
Maximization of Forcing Model, \textit{at phase extrema}
\begin{align}\label{eq:forcedmax}
\begin{split}
\mathbf[I] \: \: \: \:& \mathbb{E}(y(t)) = \frac{\mu_{{t}_A}}{t_A} \cdot t\\
\mathbf[II] \: \: \: \:& \mathbb{E}(y(t)) = \frac{\mu_{{t}_A} t_S - \mu_{{t}_S} t_A}{t_S - t_A} + \frac{\mu_{t_S} - \mu_{t_A}}{t_S-t_A} \cdot t\\
\mathbf[III] \: \: \: \:& \mathbb{E}(y(t)) = \frac{\mu_{{t}_S} t_D - \mu_{{t}_D} t_S}{t_D - t_S} + \frac{\mu_{t_D} - \mu_{t_S}}{t_D-t_S} \cdot t\\
\mathbf[IV] \: \: \: \:& \mathbb{E}(y(t)) = \frac{\mu_{{t}_D} t_R}{t_R - t_D} - \frac{\mu_{t_D}}{t_R-t_D} \cdot t\\
\end{split}
\end{align}\caption{Maximization scheme for Forcing model. Maximization of the expected utility for any listener, audience-segment-group-($i \in j$)-wise is equivalent to maximizing the probability of listening within segment, \textit{which is equivalent to maximizing each of these equations at their rightmost point}. As the mean value function for each phase has a constant first derivative the maximal path $\mathbf{x}$ is constant within phase. The budget across channels at a time $t$ is constrained by the total budget available at $t$. Again we assume that marketing spend and social buzz, etc., can only increment positively.}\label{model2max}
\end{figure}	

\begin{figure}[htbp]
\centering
 \begin{subfigure}[t]{\textwidth}
 \centering
\includegraphics[height=3cm, width=7cm]{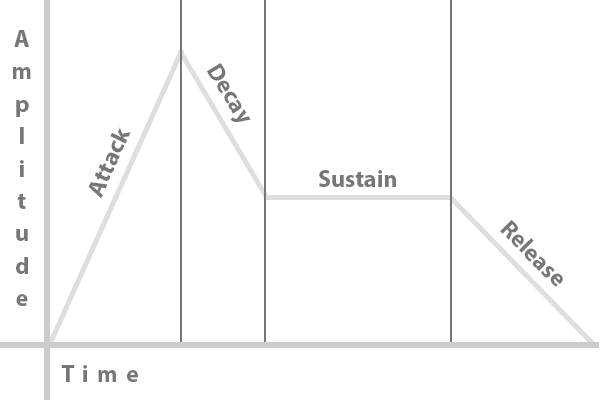} 
        \caption{ADSR model for individual sound} \label{fig:4a}
         \end{subfigure}
    \hfill
  %  \begin{subfigure}[t]{0.45\textwidth}
     %   \centering
        %\includegraphics[height=2cm]{adsr_2.jpeg} 
       % \caption{`Tuning knobs' for ADSR modeled sound} \label{fig:4b}
         % \end{subfigure}
    \vspace{1cm}
    \begin{subfigure}[t]{\textwidth}
    \centering
        \includegraphics[height=4.5cm]{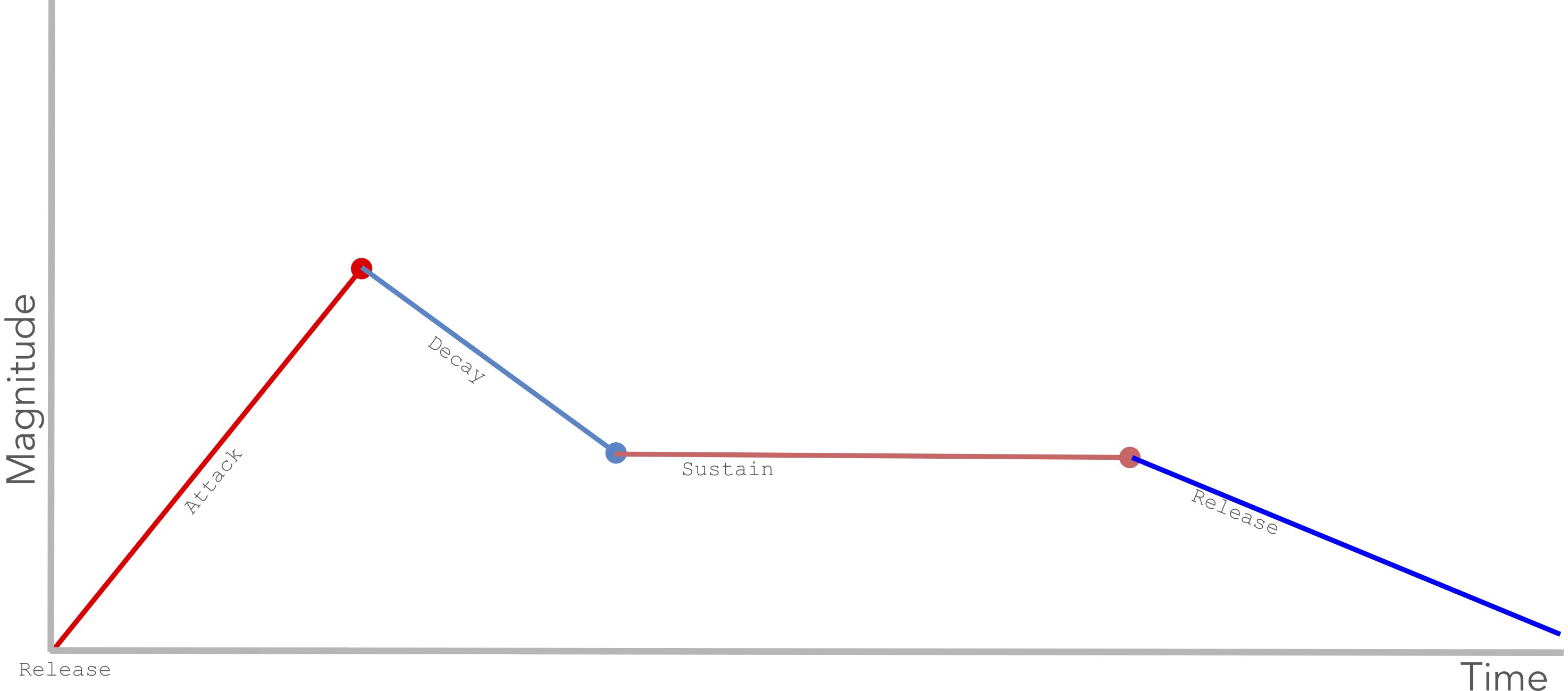} 
        \caption{ADSR model for aggregate listening demand} \label{fig:4c}
    \end{subfigure}
    \caption{Comparative illustrations of processes for song demand over time. In Figure (a\subref{fig:4a}) the model is illustrated as typically used in a Digital Audio Workstations (DAW). In Figure (\subref{fig:4c}) the model is applied to the \textit{`in-flight'} for a \textit{de novo} song from release date. This is a special case of a \textit{phase-transition model} \cite{Gomez2018}; the discontinuities here (at the nodes with enlarged circles) are where we fit partite models for each phase.}
\end{figure}

%\begin{align}
%\mathbf{x}^*_{t, i \in j} = \begin{cases} B_t[\boldsymbol\theta^j]^{-1} \; \; where \; \; 0 < B_t \leq (\mathbf{1} - \boldsymbol\gamma^j \mathbf{z})[\boldsymbol\theta^j]^{-1} \\
%(1-\boldsymbol\gamma^j \mathbf{z}^t)[\boldsymbol\theta^j]^{-1}  \; \; where \; \; B_t > (1- \boldsymbol\gamma^j \mathbf{z})[\boldsymbol\theta^j]^{-1}
%\end{cases}\label{forcedmax}
%\end{align}

\section{Comments and Recommendations}
Either of these models should `fit' nicely within current rights holder management schemes. Either model can be dynamically instantiated - in particular the phased/forcing model - with a simple LP. The forcing model needs only (linear) estimators for the mean value function at the change of phase after the change points themselves are estimated. Knowledge of these estimators - especially for this model - make a straightforward optimal path for listening maximization.

Time scales for marketing in aural media are discrete. Typically song performance is evaluated from week-to-week; advertising \& social campaigns can be adjusted weekly. Optimization schemes work well on a portfolio of assets. Use of either version of these models on a suite of assets is preferable. It is conceivable that estimators for marketing or ambient effects on listening affinity trade or switch magnitude and sign across time periods, e.g. Halloween music, Christmas music. 

An innovation shared by both the null and forcing models is to simply be willing to segregate the sources of (listening) demand and keep track of the marketing actions within each segment to yield usable time-aware effect estimators. Zooming out: audience segmentation for listening demand is key, perhaps even more for sound media demand than visual. It is not much to measure differential effects of marketing \& exposure to a sound once it is observed the same song is listened to by different audiences in different ways at different times, etc.

\section{Acknowledgment.}
The author wishes to thank his colleagues at Seton Hall University and the many hardworking data and sound scientist friends he made at Warner Music Group \& Warner Media, in particular Daniel Lee who discovered the envelope model for this use and Julien DeMori for his guidance on sound. An additional thanks to Yifeng Yu of the Music Information Program (led by Alexander Lerch) whose collaboration appears in a sequel to this paper. The author dedicates this paper to his father, Atiim Abayomi, who shared his love of music openly and who is sorely missed.

%\begin{biog}
%\item[Kobi Abayomi] received his Ph.D. in Probability \& Statistics from Columbia University. He has held postdoctoral positions at Duke University, the Statistical \& Applied Mathematical Sciences Institute (SAMSI) and Stanford University before joining the faculty of Industrial \& Systems Engineering at the Georgia Institute of Technology. Dr. Abayomi continued academic appointments (and research in Environmental and Econometric Statistics) at Binghamton University \& the University of Cuenca. Since 2015 he has led data science teams at Dun \& Bradstreet, Barnes \& Noble Education, Warner Media and most recently Warner Music Group in what he has coined \textit{Data Science for Digital Media}. Dr. Abayomi currently holds an appointment at Seton Hall University and is the Head of Science for Gumbel Demand Acceleration/Betaside Recording.
%\end{biog}
%\begin{affil}
%Department of Mathematics \& Computer Science, Seton Hall University, South Orange NJ 07040\\
%kobi.abayomi@shu.edu
%\end{affil}

%\item[Herbert Hoover] entered Stanford University in 1891, after failing all of the entrance exams except mathematics.  He received his B.S. degree in geology in 1895, spent time as a mining engineer, then was appointed by his co-author to the U.S. Food Administration and the Supreme Economic Council, where he orchestrated the greatest famine relief efforts of all time.
%\begin{affil}
%Hoover Institution, Stanford University, Stanford CA 94305\\
%herbhoover@stanford.edu
%\end{affil}
%\end{biog}
\vfill\eject

\end{document}